\documentclass[aps,prb,twocolumn,superscriptaddress, show pacs]{revtex4-1}
\usepackage{bm, amsmath}
\usepackage{graphicx}
\usepackage{color}
\usepackage{epstopdf}
\usepackage{graphicx}
\usepackage{dcolumn}
\usepackage{bm}
\usepackage{subfigure}

\begin{document}

\title{Surface twist instabilities and skyrmion states in chiral ferromagnets}

\author{S.A. Meynell}
\author{M. N. Wilson}
\affiliation{Department of Physics and Atmospheric Science, Dalhousie University, Halifax, Nova Scotia, Canada B3H 3J5}

\author{H. Fritzsche}
\affiliation{Atomic Energy of Canada Limited, Chalk River, Ontario, Canada K0J IJ0}

\author{A.~N.~Bogdanov} \thanks{a.bogdanov@ifw-dresden.de }
\affiliation{IFW Dresden, Postfach 270016, D-01171 Dresden, Germany}

\author{T.~L.~Monchesky} \thanks{tmonches@dal.ca}
\affiliation{Department of Physics and Atmospheric Science, Dalhousie University, Halifax, Nova Scotia, Canada B3H 3J5}

\date{\today}

\begin{abstract}
{ 
In epitaxial MnSi/Si(111) films, the in-plane magnetization
saturation is never reached due to the formation of specific surface chiral modulations with the propagation direction perpendicular to the film surfaces [Wilson et al. Phys. Rev. B 88, 214420 (2013)].
In this paper we show that the occurrence of such \textit{chiral surface twists} is  
a general effect attributed to all bulk and confined magnetic crystals lacking inversion symmetry.
We present experimental investigations of this phenomenon in nanolayers of MnSi/Si(111)
supported by detailed theoretical analysis within the standard phenomenological model.
In magnetic nanolayers with intrinsic or induced chirality, such surface induced instabilities 
become sizeable effects and play a crucial role in the formation of skyrmion lattices and 
other nontrivial chiral modulations.

}
\end{abstract}


\maketitle

\vspace{5mm}

\section{Introduction}

 In noncentrosymmetric magnetic crystals,
chiral asymmetry of the exchange coupling that arises due to the
crystallographic handedness leads to the formation of
long-period modulations of the magnetization
with a fixed rotation sense \cite{Dzyaloshinskii:1964jetp}.
Phenomenologically, these \textit{chiral} Dzyaloshinskii-Moriya (DM) 
interactions are described by energy contributions linear in the
first spatial derivatives of the magnetization \cite{Dzyaloshinskii:1964jetp}
\begin{equation}
\mathcal{L}_{ij}^{(k)}= m_i\partial_k m_j - m_j\partial_k m_i,
\label{form}
\end{equation}
where $\mathbf{m}$  is a unit vector along the magnetization $\mathbf{M}$,
$M = |\mathbf{M}|$ and $\partial_k m_i \equiv \partial m_i/\partial x_k$.

Theoretical analysis shows that invariants of type (\ref{form})
stabilize  different types of \textit{homochiral}  modulations that
propagate in one direction (\textit{helices}) and two directions
(\textit{skyrmion lattices}) over a broad range of thermodynamical parameters
 \cite{Dzyaloshinskii:1964jetp,Bogdanov:1989jetp,Bogdanov:1994jmmm}.
During the last four decades, homochiral long-period modulations 
have been reported in many of noncentrosymmetric  
ferro- and antiferromagnets, although only in a form of simple helical waves
 (see bibliography in Ref.~\onlinecite{Izyumov:1984spu,Bogdanov:2002prb,Rossler:2011jpcs}).
In bulk chiral magnets complex multidimensional chiral modulations including specific skyrmionic textures are reported to exist only in close vicinity to the Neel temperatures of MnSi and other cubic helimagnets as so called \textit{precursor states} (e.g., Ref.~\onlinecite{Muhlbauer:2009sci,Pappas:2009prl,Wilhelm:2011prl,Wilhelm:2012jpcm,Moskvin:2013prl,Seki:2012prb1} and bibliography in Ref.~\onlinecite{Wilhelm:2012jpcm}).
The synthesis of nanolayers of cubic helimagnets 
\cite{Yu:2010nat,Karhu:2010prb} and ultra-thin films of common magnetic metals with induced DM interactions \cite{Bode:2006nm} radically changed the situation. During last few years a large variety of skyrmion-lattice states 
and nontrivial helical modulations have been discovered in free-standing films and 
cubic helimagnetic epilayers  \cite{Yu:2010nat,Wilson:2012prb,Huang:2012prl,Wilson:2013prb}
and Fe nanolayers \cite{Heinze:2011np,Romming:2013sci}.

A sufficiently strong magnetocrystalline anisotropy
or/and an applied magnetic field will destroy chiral modulated states and establish homogeneously
magnetized phases as thermodynamically stable states 
\cite{Dzyaloshinskii:1964jetp,Bogdanov:1994jmmm}.
In such ``saturated'' phases,  the DM interactions stabilize two- and three-dimensional 
localized states (\textit{chiral} skyrmions) \cite{Bogdanov:1989jetp}.
Solutions for axisymmetric 2D chiral  skyrmions have been investigated in a number of papers
\cite{Bogdanov:1994jmmm,Butenko:2010prb,Kiselev:2011jpd,Rybakov:2013prb,Wilson:2014prb}, numerical solutions for 3D isolated chiral skyrmions  
 (also known as \textit{hopfions}) are derived in Ref.~\onlinecite{Rybakov:2010fnt}.
Isolated chiral 2D skyrmions have been recently observed in cubic helimagnet films \cite{Yu:2010nat}
and FePd nanolayers \cite{Romming:2013sci}. These axisymmetric solitonic states of nanoscale size  
have attracted a lot of attention both in fundamental nonlinear physics and as promising objects 
for different spintronic applications \cite{Kiselev:2011jpd,Romming:2013sci,Fert:2013nn}. 

It was found recently  that the solutions for skyrmions and helicoids 
in thin layers of cubic helimagnets are inhomogeneous along the film thickness:
their structure can be thought as a superposition of common in-plane modulations and
specific \textit{twisted} modulations propagating in the perpendicular direction
 \cite{Rybakov:2013prb}. 
These twisted distortions drastically change the energetics of skyrmion and helicoid 
phases and stabilize them in a broad range of applied magnetic fields. 
Experimental investigations of MnSi/Si(111) epilayers in strong in-plane magnetic 
fields have led to the discovery of twisted chiral modulations in a form of localized 
surface states \cite{Wilson:2013prb}. These observations and supporting theoretical analysis  
show that the surface twists persist at arbitrarily high fields and prevent a total saturation of the magnetization.  Evidence for this type of surface modulation can also be found in calculations of chiral modulations in confined cubic and uniaxial 
helimagnets  \cite{Du:2013epl,Rohart:2013prb,Iwasaki:2013nn,Sampaio:2013nn}.

In this paper we present experimental and theoretical investigations of
surface modulation instabilities (twisted states) that arise in MnSi/Si(111) epilayers.
We show that the occurrence of such surface chiral twists is a general
phenomenon that exists in both bulk and confined noncentrosymmetric magnets.
We develop a consistent theory of surface twists and investigate their 
influence on modulated phases in chiral magnets.

\section{Chiral surface twists in confined noncentrosymmetric ferromagnets}

\subsection{Model}

Within the phenomenological theory introduced by Dzyaloshinskii \cite{Dzyaloshinskii:1964jetp}, 
the  magnetic energy density of a  non-centrosymmetric ferromagnet 
can be written as \cite{Dzyaloshinskii:1964jetp,Bogdanov:1994jmmm}
\begin{equation}
w =A\,(\mathbf{grad}\,\mathbf{m})^2 +w_{0} (\mathbf{m}) + w_D\, (\mathbf{m}) ,
\label{density}
\end{equation}
where $A$ is the exchange stiffness constant, and $\mathbf{m} = (\sin\theta\cos\psi,$ $\sin\theta\sin\psi,\cos\theta)$.

The energy density $w_{0} (\mathbf{m})$ consists of
magnetic interactions independent of 
spatial derivatives of the order parameter
and includes Zeeman and magnetocrystalline anisotropy ($w_a$)
energies as main contributions,
$w_0  = -\mathbf{H} \cdot \mathbf{m}\,M + w_a$
($\mathbf{H}$ is the external magnetic field).
The Dzyaloshinskii-Moriya (DM) energy functional $w_D$ contains the so
called \textit{Lifshitz invariants}, which are a
combination of the differential forms (\ref{form})
compatible with the crystal symmetry \cite{Dzyaloshinskii:1964jetp, Bogdanov:1989jetp}.
Particularly, for noncentrosymmetric uniaxial ferromagnets belonging 
to crystal classes $(nmm)$ and $(n22)$ ($n$ = 3, 4, 6),
such as multiferroics B$_{1-x}$A$_x$FeO$_3$ (A = Ca, Sr, Pb, Ba) with
space group $R3c$ \cite{Khomchenko:2008jap} and helimagnets
 CsCuCl$_3$ (P6$_1$22) \cite{Bugel:2001prb} and 
Cr$_{1/3}$ Nb S$_2$ (P6$_3$22) \cite{Togawa:2012prl, Ghimire:2012prb}, the DM energy functionals have the following form:\cite{Bogdanov:1989jetp}
\begin{eqnarray}
(nmm) &:& \quad w_D=D(\mathcal{L}_{xz}^{(x)}-\mathcal{L}_{yz}^{(y)});\label{Lifshitz1} \quad\\ 
(n22) &:& \quad w_D = D(\mathcal{L}_{xz}^{(y)}
+\mathcal{L}_{zy}^{(x)}) + D_1 \mathcal{L}_{yx}^{(z)}.\label{Lifshitz2} \quad 
\end{eqnarray}
When the two Dzyaloshinskii constants are equal, $D = D_1$, the functional $w_D$ (\ref{Lifshitz2}) becomes,
\begin{eqnarray}
 w_D = D(\mathcal{L}_{yx}^{(z)}+\mathcal{L}_{xz}^{(y)}
+\mathcal{L}_{zy}^{(x)}) \equiv D\mathbf{m}\cdot \rm{rot} \mathbf{m},
\label{mrotm}
\end{eqnarray}
which describes the DM energy of cubic helimagnets with a B20-structure (space group P2$_1$3) 
such as MnSi and FeGe \cite{Bak:1980jpc}.

The equilibrium states of a confined noncentrosymmetric ferromagnet are derived by minimization of
the energy functional 
\begin{eqnarray}
 W =  \int_V w (\mathbf{m}) d\mathbf{r}+ \int_{\partial V} W_s (\mathbf{m}) d\mathbf{r}
\label{energy}
\end{eqnarray}
with corresponding boundary conditions \cite{Akhiezer:1968}.
The first integral over the  sample volume includes the energy contributions (\ref{density}), 
and the second integral over the surface comprises surface-related energies with a surface
density $W_s (\mathbf{m})$.

\subsection{1-D Isolated and bound surface twists} \label{sec:1Dtheory}

\begin{figure}
\includegraphics[width=1.0\columnwidth]{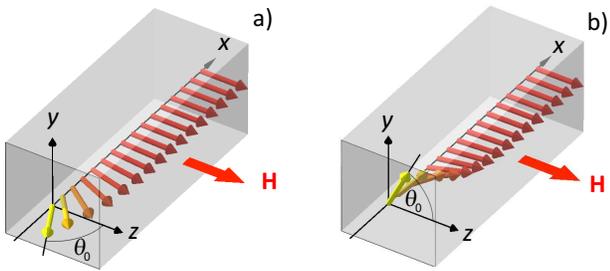}
\caption{
(color online). Localized chiral twists created
near the lateral face $(x=0)$ of a uniaxial noncentrosymmetric
ferromagnets with $nmm$ ($C_{nv}$) symmetry (a), and a cubic helimagnet or a uniaxial ferromagnets with $n22$ ($D_n$) symmetry (b), in applied fields $H>H_D$.
The angle $\theta_0$ is the angles of magnetization with respect to the field direction at the surface of the magnet.
\label{twist0}
}
\end{figure}

Here we consider a bulk noncentrosymmetric ferromagnet with chiral modulations suppressed by an applied magnetic field or a uniaxial magnetocrystalline anisotropy.  
 We show that Lifshitz invariants $\mathcal{L}_{ij}^{(k)}$ (Eq.~(\ref{form})) create surface modulations in an otherwise uniform magnetic state.
As a model we consider a semi-infinite chiral ferromagnet that occupies $x \geq 0$ 
with an applied magnetic field along the $z-$axis.  We first discuss 1-D modulations where the angle of the magnetization with respect to the $z-$axis, $\theta(x)$, remains homogeneous in the $yz$ planes.  The magnetization twists about the $x > 0$ semiaxis from $\theta (0) = \theta_0$ at the surface to $\theta(\infty) = 0$  in the depth of the sample.
Depending on the form of the Lifshitz invariants, the modulations will either be cycloidal,
$ \mathbf{m} = (\sin \theta, 0, \cos \theta)$
for uniaxial ferromagnets with $(nmm)$ symmetry (Eq.~\ref{Lifshitz1}) or helicoidal, 
$\mathbf{m} = (0, \sin \theta, \cos \theta)$
for uniaxial ferromagnets with $(n22)$ symmetry 
(Eq.~\ref{Lifshitz2})
and for cubic helimagnets
as shown in Fig.~\ref{twist0}(a) and (b) respectively.

The energy functional for isolated one-dimensional surface modulations can then be written as,
\begin{eqnarray}
 W = \int_{0}^{\infty} \underbrace{\left[ A \,\theta_x^2 
- D \,\theta_x +g (\theta) \right] }_{w(\theta, \theta_x)}dx
+W^{(s)} (\theta)|_{x = 0},
\label{energy2}
\end{eqnarray}
with the boundary conditions $\theta (0) = \theta_0 $,  
$\theta (\infty) = 0 $ and $\theta_x (\infty) = 0 $ ( $\theta_x \equiv d \theta / d x$).
The potential $g(\theta)$ represents the energy contributions that 
do not depend on spatial derivatives, such as bulk anisotropies
and is defined as  
$g (\theta) = w_0 (\theta) - w_0 ( 0)$, the difference between
energy densities of the modulated state ($w_0 (\theta)$) and that of the homogeneous
state $ w_0 (0)$. 
The boundary energy is given by $W^{(s)} (\theta)$.

\begin{figure}
\includegraphics[width=1.0\columnwidth]{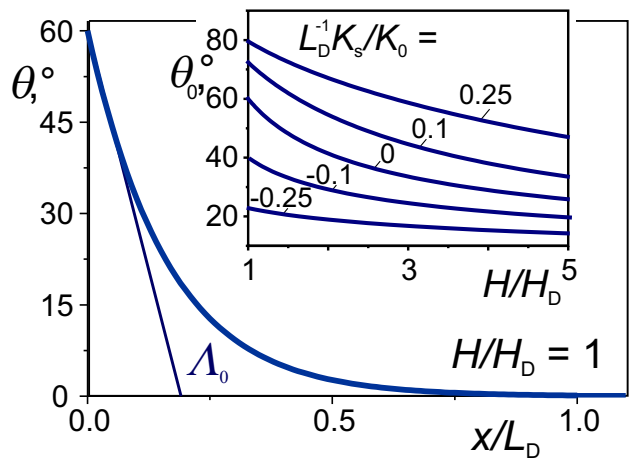}
\caption{
(color online). Magnetization profile 
$\theta (x)$  describes localized chiral twists
near lateral surfaces of a isotropic helimagnet 
at in-plane magnetic field $H = H_D$
(Eq. (\ref{sol1})) . Inset shows $\theta_0$ as a function
of the applied field for different values of uniaxial surface
anisotropy ($K_s$) (solutions to Eq. (\ref{theta02})).
\label{twist}
}
\end{figure}

First we consider solutions in a cubic helimagnet that neglect 
magnetic anisotropies, which are typically weak in these systems.
In a bulk (infinite) cubic helimagnet, the DM interaction induces
single harmonic chiral modulations propagating along the applied
magnetic field  with wave vector $k_0 = 2\pi/L_D$
(the \textit{cone} phase) \cite{Bak:1980jpc}. The helix period, 
$L_D$ and saturation field, $H_D$ are expressed via basic
magnetic parameters, \cite{Bak:1980jpc} 
\begin{eqnarray}
 L_D = 4 \pi A /|D|, \quad H_D = D^2/(2 A M).
\label{period}
\end{eqnarray}
For $H > H_D$, a helimagnet is in the saturated state with $\mathbf{m} || \mathbf{H}$.
In this case $g(\theta) = HM (1 - \cos \theta)$.
Minimization of functional (\ref{energy2}) with $W_s = 0$ (free boundary conditions) and a subsequent optimization of the mean energy with respect
of parameter $\theta_0$ yield the following analytical solutions for the chiral surface states
in a cubic helimagnet
\begin{eqnarray}
\theta(x) = 4 \arctan \left[ C(H) \exp \left(-2 \pi x \sqrt{H/H_D}/L_D \right) \right], \ \ \
\label{sol1}
\end{eqnarray}
where $C (H) = 2 \sqrt{H/H_D} - \sqrt{4(H/H_D)-1}$ and
\begin{eqnarray}
\theta_0 = 2 \arcsin \sqrt{H_D/(4H)} . 
\label{sol2}
\end{eqnarray}
For positive  DM constants ($D > 0$) in functional (\ref{energy2}), twisted states
have clockwise modulations, and an opposite rotation sense for
$D < 0$ (Fig. \ref{twist}).

The spatial extent of these surface states can be characterized by 
a \textit{penetration length} defined as the point where the
tangent at $x = 0$ intersects the $x$-axis (Fig. \ref{twist}),

\begin{eqnarray}
\Lambda_0 = -\theta_0 \left(  d \theta/ dx \right)^{-1}_{x = 0} 
= \theta_0 L_D /(2\pi) . 
\label{lambda}
\end{eqnarray}

Figure~\ref{twist} shows a typical distribution of the magnetization 
in surface chiral twist states, as given by
Eq. (\ref{sol1}).  At $H = H_D$ 
the deviation of the magnetization on the surface reaches
value $\theta_0 = 60^\circ$ at $H = H_D$
and slowly decreases  in increasing
magnetic field (Inset in Fig. \ref{twist})).

In a general case of functional (\ref{energy2})
the Euler equation yields the first integral and the solution for 
a penetration length,
\begin{eqnarray}
 A \,\theta_x^2 - g (\theta) = 0, \quad
\Lambda_0 = \theta_0 \sqrt{A/g(\theta_0)}.
\label{sol0}
\end{eqnarray}
Inserting solutions (\ref{sol0}) into (\ref{energy2})
leads to the following expression for
the equilibrium energy 
\begin{eqnarray}
 \bar{W} (\theta_0) = 2 \sqrt{A} \int_{0}^{\theta_{0}} 
\sqrt{g(\theta)} d \theta  -D \theta_0 + W_s (\theta_0) 
\label{energy3}
\end{eqnarray}
Optimizing (\ref{energy3}) with respect to $\theta_0$
leads to the equation 
\begin{eqnarray}
2 \sqrt{A g (\theta_0)} - D + d W_s (\theta_0)/d \theta_0 = 0.
\label{theta0}
\end{eqnarray}
For cycloidal modulations, there is a demagnetizing field that takes the form of a uniaxial anisotropy that must be included in the potential $g(\theta)$, whereas the helicoidal modulations in this geometry have no stray magnetic field. 
For the case of helicoidal modulations and when $W_s = 0$ Eqs. (\ref{sol0}), (\ref{theta0}) yield the solutions
\begin{eqnarray}
\theta_x (0) = 2 \pi/L_D, \quad  2 g (\theta_0) = M H_D.
\label{free1}
\end{eqnarray}
Note that in this case $\theta_x (0)$ is equal to the wave vector
 of helical modulations propagating in bulk cubic helimagnets, $k_0$
(\ref{period}). Finite values of spatial derivatives of the magnetization
on the surface of saturated chiral ferromagnets signify their instabilities
with respect to chiral surface modulations. 
By using the first integral of functional (\ref{energy2}), Eq. (\ref{theta0})
can be transformed into the following form
\begin{eqnarray}
\left[2 A \theta_x + D - d W_s (\theta)/d \theta \right]_{x =0} = 0,
\label{boundary}
\end{eqnarray}
which coincides with the regular boundary condition 
$ d (\partial w /\partial \theta_x) / dx $ =
$-\partial W_s / \partial \theta$ for functional (\ref{energy2}). The influence
of boundary condition (\ref{boundary}) on vortex and skyrmion states in magnetic
nanodots has been investigated in Refs.~\onlinecite{Du:2013epl,Rohart:2013prb,Sampaio:2013nn}.

Chiral symmetry breaking and a modification of electronic and chemical properties 
at surfaces and interfaces of magnetic nanolayers induce specific magnetic interactions
responsible for a number of phenomena unknown in bulk counterparts of the same
magnetic materials \cite{Heinrich:1993ap}. These induced interactions either penetrate into the depth of nanolayers and are described by different forms of volume energy contributions or are localized
near the surface and described by surface energy contributions 
 \cite{Heinrich:1993ap, deJonge:1994fk, Bogdanov:2002jmmm}.
The latter usually include surface/interface uniaxial anisotropy energy,  $W_{a} ^{(s)}$,
and/or the energy of exchange coupling with adjacent magnetic layers 
(so called unidirectional anisotropy),  $W_{u}^{(s)}$
\begin{eqnarray}
W_{a} ^{(s)}= K_s \cos^2 \theta, \quad W_{u}^{(s)} = K_{u} \cos \theta.
\label{surfaceenergy}
\end{eqnarray}

In chiral magnets, surface contributions of type (\ref{surfaceenergy})
influence the parameters of chiral twists. Particularly, for
$W_s = K_s \cos^2 \theta$,   Eq.~(\ref{theta0}) for the equilibrium
$\theta_0$ reads
\begin{eqnarray}
2 \sqrt{\frac{H}{H_D}} \sin \frac{\theta_0}{2} - 1
 - \left( \frac{K_s}{K_0} \right) \frac{2\pi}{L_D} \sin \theta_0 \cos \theta_0 = 0, \ \ \ \
\label{theta02}
\end{eqnarray} 
where $K_0 = D^2 / (4 A)$.
The solutions of Eq.~(\ref{theta02}) plotted as functions $\theta_0 (H/H_D)$  
for different values of $K_s$ (Inset of Fig. \ref{twist})
show that easy-plane anisotropy ($K_s > 0$) extends the twisted modulations while
easy-axis anisotropy ($K_s < 0$) suppresses them. Note that $\theta_0$ 
preserves finite values even in a case of strong easy-axis surface anisotropy
($\theta_0 = [ \sqrt{H/H_D} + 2 \pi L_D^{-1} (K_s/K_0) ]^{-1} << 1$).

Twisted states also arise in the more common achiral magnets
and other condensed matter systems as a result of a competition 
between surface and volume interactions.
Initially, twisted states have been identified
in thin layers of nematics 
confined between two parallel plates
(\textit{Freedericksz} transition
\cite{deGennes:1993}). 
Later, similar surface-induced distortions
have been observed in many classes of liquid crystals, 
and now this phenomenon forms the physical basis 
for liquid-crystal display technologies \cite{deGennes:1993}.
Theoretical investigations of magnetic twisted states
have been started by Goto et al. \cite{Goto:1965jap}
and have been developed in a number of other
works (see e.g. \onlinecite{Thiaville:1992jmmm,Bogdanov:2002jmmm,
Bogdanov:2003prb}) 
including twists in the magnetization near the surface of bulk ferromagnets \cite{Mills:1989prb,OHandley:1990prb}.
Twisted states in centrosymmetric confined ferro- and antiferromagnets 
have an arbitrary rotation sense and arise only under the condition of
specific relations between surface and volumes magnetic
anisotropies \cite{OHandley:1990prb,Thiaville:1992jmmm,Bogdanov:2003prb}.
Currently, there is only indirect experimental evidence 
for the existence of these surface states in centrosymmetric 
magnets.\cite{Tang:1993prl, Kawauchi:2011prb}

\subsection{Twisted distortions of skyrmions and helicoids}

In the depth of bulk noncentrosymmetric ferromagnets skyrmions are homogeneous along their axes ($z$-axis
in this paper) and  are described by axisymmetric solutions of type  $\theta (\rho)$, $\psi (\varphi )$
\cite{Bogdanov:1989jetp,Bogdanov:1994jmmm} (particularly, $\psi = \pi/2 + \varphi $ in cubic
helimagnets and in uniaxial ferromagnets with $n22$ symmetry).
Near the sample surfaces Lifshitz invariants with gradients along 
the skyrmion axis impose twisted distortions.
In cubic helimagnets these distortions can be written as
$\theta (\rho, z)$, 
$\psi  (\varphi,z)  = \frac{\pi}{2} + \varphi + \tilde{\psi}(z)$,
and are derived by minization of energy functional 
\begin{eqnarray}
&&w = A \left( \theta_z^2 + \theta_{\rho}^2 + \frac{\sin^2 \theta}{\rho^2}  \right)
- D \left(  \theta_{\rho} + \frac{ \sin 2 \theta}{2 \rho} \right) \cos \tilde{\psi}(z)
\nonumber \\
&&
+\sin^2 \theta \left( A \psi_{z}^2 - D \psi_z \right)  +2 MH \sin^2 \frac{\theta}{2}
\label{skyrmion0}
\end{eqnarray} 
In common magnetically soft materials, magneto-dipole interactions play an important role \cite{Hubert:1998}. However, 
the effects of spatially inhomogeneous stray-fields in noncentrosymmetric magnets
are weakened by the DM interactions \cite{Kiselev:2011jpd}, and are ignored here.

In bulk helimagnets and thick layers, with thickness $d$ much larger than
the penetration depth $d \gg \Lambda_0$ given by Eq.~(\ref{lambda}), the twisted modulations exist only
in a narrow vicinity of the surface, and can be considered as localized states.
The expansion of the skyrmion energy leads to the  
energy functional $\widetilde{W} = \int_0^{\infty} \tilde{w} (\psi, z) dz$ for the twisted states $\tilde{\psi} (z)$, where 
\begin{eqnarray}
\tilde{w} (\tilde{\psi} , z) =  \Omega_1 \left( A \tilde{\psi} _z^2 - D \tilde{\psi} _z \right) + 2D \Omega_2 \sin^2 (\tilde{\psi} /2),
\label{skyrmiontwist1}
\end{eqnarray} 
$ \Omega_1 = \int_0^{\infty} \rho d \rho \sin^2 \tilde{\theta}$,
$ \Omega_2 = \int_0^{\infty} \rho d \rho [\tilde{\theta}_{\rho}+\sin 2\tilde{\theta} /(2\rho)]$ and
$\tilde{\theta} (\rho)$ are the solutions for skyrmions in the depth of the sample, where $\tilde{\theta}_\rho \equiv d \tilde{\theta} / d \rho$.
Energy density (\ref{skyrmiontwist1}) has the same functional form as that for isolated one-dimensional
twists, Eq. (\ref{energy}), and thus surface twists are described by solutions of type (\ref{sol1}).
Similar chiral surface twists arise in helicoids.
In a thin helimagnet layer, with $d$ of the order of $\Lambda_0$, twisted modulations exist across the depth of the film. The solutions for skyrmions and helicoids in thin cubic helimagnet films have
been investigated in Ref. \onlinecite{Rybakov:2013prb}. It was found that twisted modulations substantially
modify the energetics of chiral modulations and stabilize helicoids and skyrmion lattices in a broad range
of applied fields \cite{Rybakov:2013prb}.

\begin{figure}
\includegraphics[width=1.0\columnwidth]{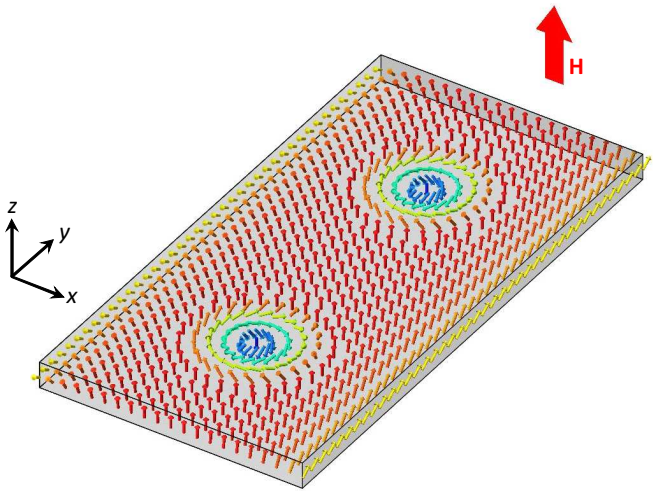}
\caption{
(color online). Repulsive chiral surface twists that arise 
at the edges of a narrow strip of a saturated helimagnet 
force isolated skyrmions to move along the central line 
of the strip.
}
\label{track}
\end{figure}

The interaction between homochiral localized modulations arising in noncentrosymmetric magnets
has a repulsive character. Consider a semi-infinite sample with the surface at $x=0$, 
and an isolated skyrmion with its axis along the $z-$axis.  When the skyrmion and 
a localized surface twist are separated by a large distance $r \gg \Lambda_0$, 
the interaction potential includes the DM energy of a surface twist 
as the main contribution to $W_{\mathrm{int}} (r) = D \int_{r/2}^{\infty} \theta_{\rho} dx = D \theta (r/2)$
\cite{Bogdanov:1995jetpl}. The expansion of Eq. (\ref{sol1}) for $ \theta \ll 1$ yields
\begin{eqnarray}
W_{\mathrm{int}} (r) =  4D C (H) \exp{ \left[-\pi \sqrt{H/H_D} \left(r/L_D \right) \right]} 
\label{interaction}
\end{eqnarray} 
Recent numerical simulations have demonstrated that chiral skyrmions are easily induced, transported and controlled by electric currents and applied magnetic fields in narrow strips of helimagnets
\cite{Sampaio:2013nn,Iwasaki:2013nn}.  
In such a track, the repulsive forces imposed by surface twists create a 
lateral confining potential for moving skyrmions along the center of the strip ($x=0$),
$w_{\mathrm{int}} (x) = [W_{\mathrm{int}} (r-x)+W_{\mathrm{int}} (r+x)]/(2 W_{\mathrm{int}} (r))$ for
$x \ll r$ (Fig. \ref{track}).

\section{Measurements of the Twisted state in MnSi thin films}
In bulk cubic helimagnets, at low temperature the magnetic structure crosses two transition fields on its way to saturation.  The first transition at a field $H_{C1}$ corresponds to a rotation of the propagation vector in the direction of the field and a canting of the spins into a conical state\cite{Plumer:1981jpc,Grigoriev:2006prb1}. This is followed by a second-order collapse of the conical phase into a field induced ferromagnetic state above a field $H_{C2}$.   
The situation for epitaxial cubic helimagnetic thin films is altered by strain induced uniaxial anisotropy and the breaking of symmetry caused by the presence of the interfaces. In the case of MnSi thin films, the ground state is a helix with an out-of-plane propagation vector\cite{Karhu:2011prb}.
At low temperatures, the uniaxial anisotropy prevents a reorientation of the helix in magnetic fields applied in-plane, and the propagation vector remains fixed along the film normal.   The in-plane field produces helicoids with a discrete number of turns, and the magnetic state evolves by a set of first-order transitions between these states.\cite{Wilson:2013prb}  The final state is not the uniform ferromagnetic state envisioned in bulk crystals, but rather is a ferromagnetic state with surface twists, as demonstrated by the polarized neutron reflectometry curves in Ref.~\onlinecite{Wilson:2013prb}, that show no signs of saturation. 

To compare with the analytical expressions in this Paper, we conducted SQUID magnetometry of a series of samples that range between 11.5 nm and 40 nm.  Samples were grown by molecular beam epitaxy on Si(111) substrates as described in Ref.~\onlinecite{Karhu:2012prb}, and measurements were conducted in a Quantum Design MPMS XL5 SQUID magnetometer.  The helical wavelength $L_D=13.9$~nm is obtained from polarized neutron reflectometry (PNR) and SQUID analysis.\cite{Karhu:2011prb}  SQUID magnetometry of MnSi/Si(111) is complicated by the large diamagnetic contribution from the Si substrate.  One typically determines the substrate contributions from the high-field portion of the $M$-$H$ loops, far above the saturation field.  However, in the case of MnSi, the high-field susceptibility is non-zero,\cite{Bloch:1975pla} which makes it impossible to separate the film's contribution from that of the substrate with this method.

In order to determine the high-field susceptibility of our MnSi films, we performed PNR on a $d=26.7$~nm MnSi film at a temperature of $T = 40$~K, below the Curie temperature of $T_C = 44$~K. 
PNR provides a depth dependent measurement of the magnetization of the film without any contribution  from the diamagnetic response of the substrate.  We conducted measurements with 0.237 nm neutrons on the D3 reflectometer at the Canadian Neutron Beam Centre mounted with the M5 superconducting magnet cryostat.\cite{Fritzsche:2005rsi}  An Fe/Si supermirror and a Mezei-type precession spin flipper produced a neutron beam with a spin polarization in excess of 95\%.
To avoid attenuating the PNR signal, we conducted the experiments without an analyzer, since our previous measurements show that the spin-flip signal is negligible due to the cancelling effects of the bi-chiral magnetic domains.\cite{Karhu:2012prb}  
From the measured spin-up $R_{+}$, and spin-down, $R_{-}$, reflectivities, we calculated the spin-asymmetery $\alpha = (R_{+} - R_{-})/(R_{+} + R_{-})$.
The first peak in $\alpha$ at low scattering vectors $0.08~\le q \le 0.17$~nm provides an excellent measure of the average magnetization across the film.  

\begin{figure}[!h]
\begin{center}
\includegraphics[width=\columnwidth]{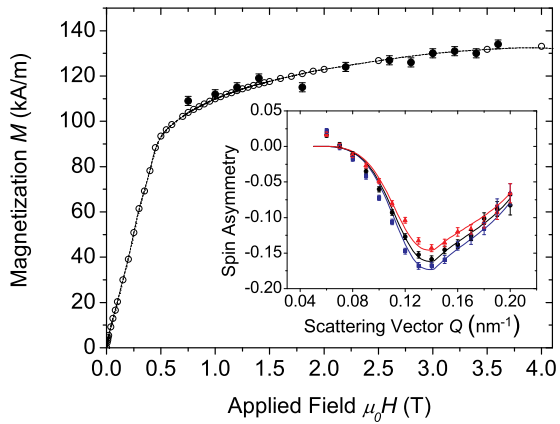}
\caption{ (color on-line) The inset shows the three representative measurements of the PNR spin-asymmetry $\alpha$ from a $d = 26.7$-nm $= 1.92 L_D$ thick MnSi film measured at $T = 40$~K in fields $\mu_0 H = 1.0$~T (red triangles), 2.2~T (black circles) and 3.2~T (blue squares).  The lines are fits to the data described in the text. The magnetization obtained from the fits is plotted in the main figure (filled points) along with the substrate-subtracted SQUID magnetometry data (open points). The line is a guide to the eye. All error bars are $\pm1 \sigma$.}
\label{fig:MH}
\end{center}
\end{figure}

In order to extract the magnetization, we fitted the measurement of $\alpha$ in the inset of Fig.~\ref{fig:MH} to simulations calculated with SimulReflec.  
The calculated non-spin-flip reflectivities, $r_{++}$ and $r_{--}$, are corrected for non-idealities in the beam polarization before they are compared with experiments.  The corrections,
\begin{eqnarray}
R_{+} &=& r_{++} + \frac{r_{--}}{2 F} \ , \\
R_{-} &=& \frac{r_{++}}{2 F} + r_{--} \ ,
\label{FRcorr}
\end{eqnarray}
also account for the absence of the analyzer during the reflectivity experiments.  The flipping ratio, $F = I^{++}/I^{+-}$, is obtained by measuring the unreflected beam intensities $I^{++}$ and $I^{+-}$ that pass through both the polarizer and analyzer, where the subscripts denote the polarization of the neutron beam after passing through the polarizer and analyzer respectively.  The flipping ratio varies between 20 to 33 over the range of fields measured. 
The structural parameters in the simulations are obtained from the x-ray reflectometry and PNR measurements presented in Ref.~\onlinecite{Karhu:2012prb}. 
The magnetization profile is calculated from Eq.~(\ref{sol1}) by accounting for the surface twists at both interfaces.
The sole fitting parameter to the PNR data in the inset of Fig.~\ref{fig:MH} is the value of the average magnetization, which is plotted for various fields in the main figure.  
The value for the substrate susceptibility subtracted from the SQUID measurements is adjusted to bring the magnetometry measurements into agreement with the PNR data, as shown by the open circles in Fig.~\ref{fig:MH}.  Since SQUID measurements show that the Si susceptibility is independent of temperature below $T = 95$~K, we were then able to accurately remove the substrate contribution for all measured temperatures.  
The high-field susceptibility of the MnSi film at $T = 40$~K is $\chi_{HF} = 4.0 \pm 0.6$~kA/m/T, which is in good agreement with $\chi_{HF}=4.4$~kA/m/T for bulk MnSi (estimated from Ref.~\onlinecite{Chattopadhyay:2009jpcm}) at the same reduced temperature.  We therefore assume that the high-field susceptibility is the same for all thicknesses presented in this paper and use this value to remove the substrate susceptibility from the other samples.
	
\begin{figure}[!h]
\begin{center}
\includegraphics[width=\columnwidth]{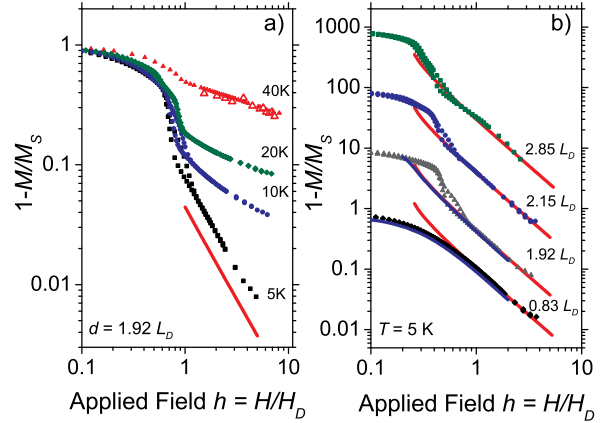}
\caption{ The reduced magnetization $\Delta m$ measured by SQUID magnetometry (filled points) for (a) $T = 40, 20, 10$ and 5~K in a $d= 1.92 L_D$ film, where $L_D = 13.9$~nm is the helical wavelength.
The estimated values for $H_D$ are 0.49, 0.75, 0.80 and 0.82~T respectively, and the high-field magnetization is $M_s = 178.7$~kA/m. 
The large open triangular points superimposed on the $T = 40$~K data correspond to PNR measurements.  The red lines in (a) and (b) are calculations given by Eq.~\ref{eq:deltam}. Figure (b) shows $\Delta m$ for four films with thicknesses from  $d = 0.83 L_D$ to 2.85~$L_D$, where the curves are vertically offset for clarity. The red lines are calculations given by Eq.~\ref{eq:deltam}, where $H_D = 1.08, 1.21, 1.38, 1.53$~T in order of increasing thickness.  The blue curves show exact numerical calculations of the twisted ferromagnetic state that account for the finite size, as described in Ref.~\onlinecite{Wilson:2013prb}.}
\label{fig:mh2}
\end{center}
\end{figure}

In the limit of large field, $H > H_D$, and large thickness, $d \gtrsim L_D$, 
the solution given by Eq.~(\ref{sol1}) provides the field dependence of the magnetization in the film,
\begin{equation}
\Delta m = \frac{2}{\pi} \frac{L_D}{d} \sqrt{\frac{H_D}{H}} 
\left( 1 - \sqrt{1 - \frac{H_D}{4 H}} \right) ,
\label{eq:deltam}
\end{equation}
where $\Delta m $ is the reduced magnetization $(1 - M/M_s)$.   To compare with theory, the saturation magnetization  $M_s(H \rightarrow \infty)$ is determined by the value of $M_s$ that gives a linear $\log \Delta m $ dependence on $\log H$ at fields $H > H_D$.  The values of $H_D$ are obtained using the procedure described in Ref.~\onlinecite{Karhu:2012prb}.  The plots of  $\log \Delta m $ vs $\log H$ in Fig. ~\ref{fig:mh2} show linear regimes for $H \gtrsim H_D$ that correspond to the twisted ferromagnetic state, and a more complex $M$-$H$ dependence for $H \lesssim H_D$ that correspond to regions with a more complex magnetic texture.   Figure~\ref{fig:mh2} (a) shows that there is a clear departure of the data from the behavior predicted by Eq.~(\ref{eq:deltam}).  However, as the temperature is dropped, the $d\log \Delta m /d\log H$ approaches the expected $H^{3/2}$ dependence at $T=5$~K.  The fact that the magnitude of the discrepancy grows with temperature suggests that spin-fluctuations may be dominating the higher temperature behavior. They would explain the change in slope of $\log \Delta m$ vs $\log H$ since the magnetization is expected to have a $H^{1/2}$ field dependence due to fluctuations.\cite{Kaul:1999jpcm}    However, even at $T=5$~K, the data in Fig.~\ref{fig:mh2} (a) is offset from the theoretical curve.  This offset can be eliminated by an increase in $H_D$.

The presence of twisted surface states required a modification to the method that we use to extract values for $H_D$, which assumes that as the field is lowered, the system transitions continuously from a saturated state into a conical state\cite{Karhu:2012prb}.  This is true for out-of-plane fields, where the saturation field,
\begin{equation}
H_{C2,\perp} = H_D + \frac{2 K_u}{\mu_0 M_s} + M_s.
\label{HC2}
\end{equation}
For in-plane fields, however, a ferromagnetic state with surface twists exists in place of a conical phase and there is no in-plane saturation field,$H_{C2,\|}$. 
If we use the minimum in $\partial^2 M/\partial H^2$, as done in Refs.~\onlinecite{Karhu:2012prb,Wilson:2012prb}, to identify an effective $H_{C2,\|}$, we obtain estimates for $H_D$ that are approximately 15\% lower than required to fit the discrete helicoidal transition fields in Ref.~\onlinecite{Wilson:2013prb}.  We use these values as crude estimates of $H_D$.

The theory for twisted states presented in Section~\ref{sec:1Dtheory} provides, in principle, a method of extracting values for $H_D$ from in-plane high magnetic field measurements of $M(H)$.  
We used Eq.~(\ref{eq:deltam}) to fit the magnetization curves by treating $H_D$ and $M_s$ as fitting parameters.  The results for films from $d = 11.5$~nm to $d = 40$~nm shown in Fig.~\ref{fig:mh2}(b) demonstrate that Eq.~(\ref{eq:deltam}) fits the data well, but unfortunately we find that the fits yield $H_D$ values that are inconsistent with previous measurements.
For example, we find $H_D = 1.4$~T for the $d=1.92 L_D$ film, which is too large given $H_{C2,\perp} = 1.18$~T and Eq.~(\ref{HC2})\cite{Karhu:2012prb}, and is inconsistent with fits to the helicoidal state for in-plane magnetic fields\cite{Wilson:2013prb}. 
Therefore the values for $H_D$ obtained by using Eq.~(\ref{eq:deltam}) can only be taken as effective characteristic fields, $H_{Deff}$, which include other factors such as surface anisotropies or extrinsic effects that affect the saturation of the twisted state. 
These results highlight the need for independent measurements of $K_u$ so that more accurate values of $H_D$ can be achieved.

In Fig.~\ref{fig:mh2} (b), we compare the analytical expression (red) with the exact numerical calculations (blue).  For the films $d < L_D$, the structure remains in a twisted ferromagnetic state down to the lowest fields, and is well described by the numerical calculation over the entire field range.  As expected, the analytical expression is in good agreement with the calculation for $H > H_D$.
In this figure, we also find good agreement between the numerical calculation and the data for the $d = 1.92 L_D$ film above $H > H_D$. The departure of the data from the model at lower fields reflects a transition to a helicoidal state at $H = 0.8 H_D$, below which the twisted state is a metastable state.
The difference between the analytical (red) and numerical (blue) curves is very small due to the fact that the twists at the two interfaces interact very little with each other at this film thickness.  For larger film thicknesses there is almost no difference, and so the numerical calculations have been left out of the figure. 

\section{Conclusions}

Experimental observations of MnSi/Si (111) cubic helimagnet epilayers  
and theoretical calculations within the standard model (\ref{density})
show that the formation of chiral surface twists is
a general phenomenon attributed to all saturated noncentrosymmetric
magnetic crystals. In bulk and confined chiral magnets for which the
thickness $d$ is much larger than \textit{penetration length} $\Lambda_0$ (\ref{sol0}),
the surface twists that arise are strongly localized near individual surfaces
and do not interact with each other.
In thin layers of chiral magnets ($ d \leq \Lambda_0$)  surface twists
in the saturated phases create bound states \cite{Wilson:2013prb}, and
helicoids and skyrmions exist as a superposition of common in-plane modulations
and conical modulations along the film thickness.
Contrary to achiral twisted phases in common (centrosymmetric) magnets,
which are predicted to exist only for specific relations between material
parameters of magnetic materials 
\cite{OHandley:1990prb,Thiaville:1992jmmm,Bogdanov:2002prb},
chiral surface twists exist in broad range of thermodynamic parameters 
and can be of practical interest
for spintronics applications \cite{Wilson:2013prb}.
\begin{acknowledgements}
We would like to thank Filipp Rybakov for helpful discussions and Eric Karhu for technical assistance.
A.N.B. acknowledges financial support by DFG through Grant No. BO 4160/1-1.
T.L.M. and M.N.W. acknowledge support from NSERC.  The research presented herein is made possible by a reflectometer jointly funded by Canada Foundation for Innovation (CFI), Ontario Innovation Trust (OIT), Ontario Research Fund (ORF), and the National Research Council Canada (NRC).
\end{acknowledgements}


%

\end{document}